\documentclass[preprint,12pt]{elsarticle}

\usepackage{epsfig}
\usepackage{here}
\usepackage{graphicx}
\usepackage{amssymb}

\usepackage{lineno}
\usepackage{comment}
\usepackage{color}
\usepackage{amsmath}

\journal{Physics Letters B}

\begin{document}

\begin{frontmatter}



\title{Search for exotic neutrino-electron interactions using solar neutrinos in XMASS-I}

\author[ICRR,IPMU]{K.~Abe}
\author[ICRR]{Y.~Chen}
\author[ICRR,IPMU]{K.~Hiraide}
\author[ICRR,IPMU]{K.~Ichimura\fnref{tohokunow}}
\author[ICRR]{S.~Imaizumi}
\author[ICRR]{N.~Kato}
\author[ICRR,IPMU]{K.~Kobayashi\fnref{wasedanow}}
\author[ICRR]{M.~Kobayashi\fnref{conow}}
\author[ICRR,IPMU]{S.~Moriyama}
\author[ICRR,IPMU]{M.~Nakahata}
\author[ICRR]{K.~Sato\fnref{nagoyanow}}
\author[ICRR,IPMU]{H.~Sekiya}
\author[ICRR]{T.~Suzuki}
\author[ICRR,IPMU]{A.~Takeda}
\author[ICRR]{S.~Tasaka}
\author[ICRR,IPMU]{M.~Yamashita\fnref{nagoyanow}}
\author[IBS2]{B.~S.~Yang}
\author[IBS1]{N.~Y.~Kim}
\author[IBS1]{Y.~D.~Kim}
\author[IBS1,KRISS]{Y.~H.~Kim}
\author[ISEE]{R.~Ishii}
\author[ISEE,KMI]{Y.~Itow}
\author[ISEE]{K.~Kanzawa}
\author[IPMU]{K.~Martens}
\author[IPMU]{A.~Mason\fnref{oxfordnow}}
\author[IPMU]{Y.~Suzuki\fnref{icrrnow}}
\author[Kobe]{K.~Miuchi}
\author[Kobe,IPMU]{Y.~Takeuchi}
\author[KRISS]{K.~B.~Lee}
\author[KRISS]{M.~K.~Lee}
\author[Miyagi]{Y.~Fukuda}
\author[Nihon,IPMU]{H.~Ogawa}
\author[RCNS,IPMU]{Y.~Kishimoto}
\author[Tokai1]{K.~Nishijima}
\author[Tokushima]{K.~Fushimi}
\author[Tsinghua,IPMU]{B.~D.~Xu}
\author[YNU1]{S.~Nakamura}

\address[ICRR]{Kamioka Observatory, Institute for Cosmic Ray Research, the University of Tokyo, Higashi-Mozumi, Kamioka, Hida, Gifu 506-1205, Japan}
\address[IBS2]{Center for Axion and Precision Physics Research, Institute for Basic Science, Daejeon 34051, South Korea }
\address[IBS1]{Center for Underground Physics, Institute for Basic Science, 70 Yuseong-daero 1689-gil, Yuseong-gu, Daejeon 305-811, South Korea}
\address[ISEE]{Institute for Space-Earth Environmental Research, Nagoya University, Nagoya, Aichi 464-8601, Japan}
\address[IPMU]{Kavli Institute for the Physics and Mathematics of the Universe (WPI), the University of Tokyo, Kashiwa, Chiba 277-8582, Japan}
\address[KMI]{Kobayashi-Maskawa Institute for the Origin of Particles and the Universe, Nagoya University, Furo-cho, Chikusa-ku, Nagoya, Aichi 464-8602, Japan}
\address[Kobe]{Department of Physics, Kobe University, Kobe, Hyogo 657-8501, Japan}
\address[KRISS]{Korea Research Institute of Standards and Science, Daejeon 305-340, South Korea}
\address[Miyagi]{Department of Physics, Miyagi University of Education, Sendai, Miyagi 980-0845, Japan}
\address[Nihon]{Department of Physics, College of Science and Technology, Nihon University, Kanda, Chiyoda-ku, Tokyo 101-8308, Japan}
\address[RCNS]{Research Center for Neutrino Science, Tohoku Univerisity, Sendai 980-8578, Japan}
\address[Tokai1]{Department of Physics, Tokai University, Hiratsuka, Kanagawa 259-1292, Japan}
\address[Tokushima]{Department of Physics, Tokushima University, 2-1 Minami Josanjimacho Tokushima city, Tokushima 770-8506, Japan}
\address[Tsinghua]{Department of Engineering Physics, Tsinghua University, Haidian District, Beijing, China 100084}
\address[YNU1]{Department of Physics, Faculty of Engineering, Yokohama National University, Yokohama, Kanagawa 240-8501, Japan}
\address{\rm\normalsize XMASS Collaboration$^*$}
\cortext[cor1]{{\it E-mail address:} xmass.publications18@km.icrr.u-tokyo.ac.jp .}

\fntext[tohokunow]{Now at Research Center for Neutrino Science, Tohoku Univerisity, Sendai 980-8578, Japan}
\fntext[wasedanow]{Now at Waseda Research Institute for Science and Engineering, Waseda University, 3-4-1 Okubo, Shinjuku, Tokyo 169-8555, Japan}
\fntext[conow]{Now at Physics Department, Columbia University, New York, NY 10027, USA}
\fntext[nagoyanow]{Now at Institute for Space-Earth Environmental Research, Nagoya University, Nagoya, Aichi 464-8601, Japan}
\fntext[oxfordnow]{Now at Department of Physics, University of Oxford, Oxford, Oxfordshire, United Kingdom}
\fntext[icrrnow]{Now at Kamioka Observatory, Institute for Cosmic Ray Research, the University of Tokyo, Higashi-Mozumi, Kamioka, Hida, Gifu 506-1205, Japan}

\begin{abstract}
  We have searched for exotic neutrino-electron interactions that could be produced by a neutrino millicharge, by a neutrino magnetic moment, or by dark photons using solar neutrinos in the XMASS-I liquid xenon detector. We observed no significant signals in 711 days of data. We obtain an upper limit for neutrino millicharge of 5.4$\times$10$^{-12} e$ at 90\% confidence level assuming all three species of neutrino have common millicharge. We also set flavor-dependent limits assuming the respective neutrino flavor is the only one carrying a millicharge, $7.3 \times 10^{-12} e$ for $\nu_e$, $1.1 \times 10^{-11} e$ for $\nu_{\mu}$, and $1.1 \times 10^{-11} e$ for $\nu_{\tau}$. These limits are the most stringent yet obtained from direct measurements. 
We also obtain an upper limit for the neutrino magnetic moment of 1.8$\times$10$^{-10}$ Bohr magnetons. In addition, we obtain upper limits for the coupling constant of dark photons in the $U(1)_{B-L}$ model of 1.3$\times$10$^{-6}$ if the dark photon mass is 1$\times 10^{-3}$ MeV$/c^{2}$, and 8.8$\times$10$^{-5}$ if it is 10 MeV$/c^{2}$. 
\end{abstract}


\end{frontmatter}

\renewcommand\thefootnote{\arabic{footnote}}

\section{Introduction}
\label{sec:intro}
Liquid xenon (LXe) detectors continue to set stringent limits on weakly interacting massive particle (WIMP) dark-matter models \cite{LUX, PANDA, XENON1T, XMASS_WIMP}. Yet these detectors are also able to explore other physics topics due to their low backgrounds (BGs) and low energy threshold.
A study using solar neutrinos was suggested in \cite{suzuki}. Solar neutrinos are generated by nuclear fusion in the Sun. The majority of solar neutrinos come from the proton-proton ($pp$) reaction, $p + p \rightarrow d + e^{+} + \nu_{e}$ in the $pp$-chain, which produces approximately 99$\%$ of the total solar energy. At Earth the flux of the $pp$ solar neutrinos is 5.98$\times$10$^{10}$ cm$^{-2}$s$^{-1}$ \cite{solarnu} and their spectrum is continuous with its endpoint at 422 keV. Another significant source of solar neutrinos is electron capture on $^{7}$Be. The flux of $^{7}$Be solar neutrinos at Earth is 5.00$\times$10$^{9}$ cm$^{-2}$s$^{-1}$ \cite{solarnu} and their energy is monochromatic 862 keV. 
Here we search for interactions between these abundant low energy solar neutrinos and the electrons in the detector's LXe target that could be signatures of a neutrino electromagnetic millicharge, a neutrino magnetic moment, or interactions mediated by dark photons.

\subsection*{Neutrino millicharge}
The electric charge of neutrinos is assumed to be  zero in the Standard Model (SM).
In general, the existence of a neutrino millicharge would give hints on models beyond the SM. In a simple extension of the SM with the introduction of the right-handed neutrino $\nu_{R}$, the neutrino is a Dirac particle and the three neutrino mass eigenstates share a common millicharge due to gauge invariance \cite{milli_3nu} whether the millicharge is zero or not. Any differences of millicharge among neutrinos and antineutrinos would be an indication of CPT violation \cite{milli_theo}. Moreover, it is still of interest to study the neutrino millicharge of each individual neutrino flavor in the unexplored parameter space.

Both, experimental searches and astrophysical indirect searches for neutrino millicharge have been performed \cite{milli_sum}, but no evidence for neutrino millicharge has been found so far. For example, the lack of a charge asymmetry in the universe constrains the neutrino charge to be $4 \times 10^{-35} e$ \cite{milli_ind}. The most stringent upper limit from direct experimental searches is $1.5 \times 10^{-12} e$ \cite{milli1}.
The limit in \cite{milli1} and the second most stringent one, $2.1 \times 10^{-12} e$ \cite{milli2}, were both obtained using reactor neutrinos, meaning electron antineutrinos, but also containing negligible amounts of other neutrino species such as $\bar{\nu}_{\mu}$ and $\bar{\nu}_{\tau}$. Thus these are antineutrino limits. 
The most stringent limit for neutrinos, on the other hand, was obtained by a vacuum birefringence experiment \cite{milli_PVLAS}. This limit is dependent on neutrino masses and $<3\times 10^{-8} e$ for neutrino masses of less than 10 meV. The limits from the reactor experiments do not have such a dependence on neutrino masses. This birefringence limit applies to all neutrino flavors. Solar neutrinos are produced as electron neutrinos, but due to neutrino oscillation at Earth they also contain $\nu_{\mu}$ and $\nu_{\tau}$. In this paper we search for millicharge in all three neutrino flavors.

\subsection*{Neutrino magnetic moment}
The massless neutrinos of the SM do not have any magnetic moment. However, a minimally-extended SM with Dirac neutrino masses predicts a finite neutrino magnetic moment of \cite{K.Fujikawa}:
\begin{equation}
\mu_{\nu} = \frac{3m_{e} G_{F}}{4\sqrt{2}\pi^{2}}m_{\nu}\mu_{B} \sim 3.2 \times 10 ^ {-19} (\frac{m_{\nu}}{1eV}) \mu_{B}
\label{eq:numag}
\end{equation}
Here $m_{e}$ is the electron mass, $G_{F}$ is the Fermi coupling constant, and $\mu_{B}$ is the Bohr magneton. Considering the observed small squared mass differences of neutrinos, the neutrino magnetic moment becomes less than $O(10^{-19}) \mu_{B}$. It is not currently feasible to detect that small a neutrino magnetic moment experimentally. However, other extensions of SM theory yield neutrino magnetic moments at currently observable levels. For example, if the neutrino is a Majorana particle, the transition magnetic moment is estimated to be $O(10^{-12}\sim10^{-10}) \mu_{B}$ in an extension that goes beyond a minimally-extended SM \cite{M.B.Voloshin}. The Borexino experiment searched for a neutrino magnetic moment using $^{7}$Be solar neutrinos. Borexino found no significant excess and set an upper limit of $2.8 \times 10^{-11} \mu_{B}$ \cite{Agostini}. Similarly, the GEMMA experiment, using reactor antineutrinos, obtained an upper limit of  $2.9 \times 10^{-11} \mu_{B}$ \cite{AGBeda}.

\subsection*{Dark photons}
There are many unsolved problems that cannot be explained by the SM, such as neutrino mass and the particle nature of dark matter, and new physics scenarios beyond the SM are required. The hidden sector scenario is one of such scenario. It could contain a dark photon, which might influence the interaction of neutrinos with electrons via dark-photon exchange. The idea that the light vector boson of this hidden sector appears as a dark photon has been around for a long time \cite{LBOkun, BHoldom}, and the possibility that it appears at low energy has received wide interest. In the context of one such scenario, we search for a dark photon derived from a gauged $U(1)_{B-L}$ symmetry, for which a noticeable increase of the cross section for electron-recoil from solar neutrino interactions is expected \cite{Harnik, Bilmis}. The mass $M_{A^{\prime}}$ of the dark photon $A^{\prime}$ and coupling constant $g_{B-L}$ are already constrained by various experimental and astrophysical analyses \cite{Bilmis}. The constraints are summarized in Figure~\ref{fig:cont_BL}. The dark photon model with $U(1)_{B-L}$ is also one of the candidates for explaining the muon $g-2$ anomaly if the dark photon mass is $O(1\sim1000)$ keV$/c^{2}$ with $g_{B-L} \sim O(10^{-4}\sim10^{-3})$ \cite{PDG}.

\vspace{0.2in}
These considerations motivate us in our search for exotic neutrino interactions. Since solar neutrinos provide the largest available flux, we used them to search for exotic neutrino interactions with the XMASS-I detector.

\section{The XMASS-I detector}
\label{sec:det}
The XMASS-I detector \cite{XMASS_detector} is located at the Kamioka Observatory in Japan, underground at a depth of 2,700 meters water-equivalent. It consists of a water-Cherenkov outer detector (OD) and a single-phase LXe inner detector (ID). The OD, which is a cylindrical water tank 11 m high and 10 m in diameter, is equipped with 72 20-inch photomultiplier tubes (PMTs) used to veto cosmic-ray muons. Data acquisition for the OD is triggered when eight or more of its PMTs register a signal within 200 ns. The ID is located at the center of the OD. An active target containing 832\,kg of LXe is held in the copper structure of the ID. The ID's inner surface is $\sim$40 cm away from the center and covered with 642 low-radioactivity PMTs (Hamamatsu R10789).  Data-acquisition is triggered for the ID when four or more hits occur within 200 ns.
Energy calibrations in the energy range between 1.2 keV and 2.6 MeV were conducted via the insertion of $^{55}$Fe,  $^{109}$Cd, $^{241}$Am, $^{57}$Co, and $^{137}$Cs sources along the vertical axis into the detector's sensitive volume, and by setting $^{60}$Co and $^{232}$Th sources outside the ID's vacuum vessel \cite{XMASS_detector, XMASS_subgevmod}. The time variation of the energy scale was traced via irradiation with $^{60}$Co every week and by the insertion of $^{57}$Co every other week.

\section{Analysis method}
\label{sec:ana}
\subsection{Simulation}
\begin{figure}[t]
  \begin{center}
    \includegraphics[keepaspectratio=true,height=55mm]{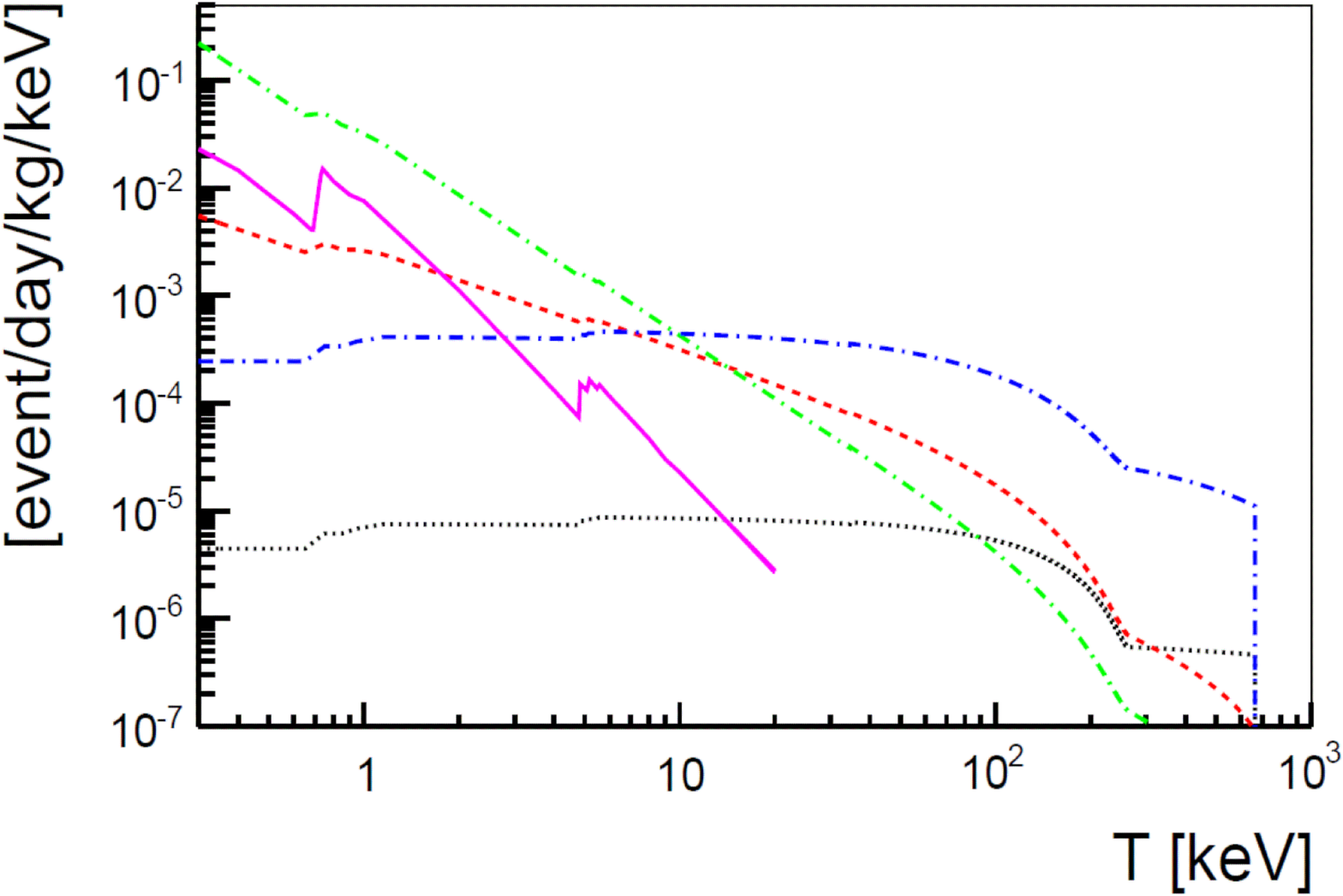}
  \end{center}
  \caption{The deposited energy spectra for neutrino interactions in xenon. The magenta-solid line shows a model where the neutrino has a millicharge (1.5$\times$10$^{-12} e$) \cite{milli_estimate}. The red-dashed line shows a model where the neutrino has a magnetic moment (1$\times$10$^{-10}\mu_{B}$) \cite{P.Vogel}. The green-dash-dotted and blue-dash-dotted line show models where neutrino interacts with electrons through dark photons with $g_{B-L}$=1$\times$10$^{-6}$ and $M_{A'}$=1$\times$10$^{-3}$ MeV$/c^{2}$ and with $g_{B-L}$=1$\times$10$^{-4}$ and $M_{A'}$=10 MeV$/c^{2}$, respectively \cite{Bilmis}. The black-dotted line shows the Standard Model neutrino-electron weak interaction. The models for atomic effects are RRPA \cite{JWChen} for millicharge and FEA \cite{VKopeikin} for magnetic moment and dark photons.}
\label{fig:exoticspec}
\end{figure}

In the process of an interaction between a neutrino and an electron mediated 
by a neutrino magnetic moment \cite{P.Vogel} or by a dark photon from the $U(1)_{B-L}$ model \cite{Bilmis}, the total number of events $N_{tot}$ is given by integrating the differential rate in free electron approximation: 
\begin{equation}
\begin{split}
\frac{dN_{\rm tot}}{dT} = &t \times N\\
&\times\int \left[\left(\frac{d\sigma_{\nu e^{-}}}{dT}\right)_{\rm SM} + \left(\frac{d\sigma_{\nu e^{-}}}{dT}\right)_{\rm ex}\right]\sum^{Z}_{i=1}\theta(T-B_{i})\left(\frac{d\Phi_{\nu}}{dE_{\nu}}\right)dE_{\nu}, 
\end{split}
\label{eq:totalrate}
\end{equation} 
where ``$\rm SM$'' indicates the term for the standard weak interaction in the SM, ``$\rm ex$'' indicates the exotic interaction term. For the dark photon analysis, interference effects with the weak interaction as in \cite{Bilmis} are included in the exotic interaction term. $T$ is the neutrino's energy deposition in the detector, which contains both the energy deposited by the recoiling electron and from subsequent transitions in the residual atom's shell, $t$ is the total livetime used in this analysis, $N$ is the number of xenon atoms, $\sigma_{\nu e^{-}}$ is the respective cross section between neutrino and electron, $E_{\nu}$ is the neutrino energy, and $\Phi_{\nu}$ is the solar neutrino flux. 
To account for atomic effects in xenon, which affect the signal expectation, we follow previous publications in using the free electron approximation (FEA) in our dark photon analysis. Effectively this approximation uses a series of step functions, one for every electron in the atom, each with the step at the respective electron's binding energy \cite{VKopeikin}. In our millicharge analysis on the other hand we follow \cite{milli_estimate} and use their results from their ab-initio multi-configuration relativistic random phase approximation (RRPA) \cite{JWChen}. At 5 keV deposited energy the FEA cross section is about a factor of five less than the RRPA one. FEA was adopted for the dark photon analysis to be consistent with the magnetic moment analysis. In the magnetic moment analysis, we used FEA because RRPA calculations were only available below 20 keV. For this energy region, FEA predicts 5$\%$ less signal than the calculation based on RRPA. Thus the results of our neutrino magnetic moment and dark photon analyses based on FEA are conservative relative to what would be expected for RRPA.
Figure~\ref{fig:exoticspec} shows the deposited energy spectra of neutrino-electron interactions in xenon. The event rate due to dark photons is proportional to the forth power of $g_{B-L}$ and the spectral shape depends upon $M_{A^{\prime}}$ while the event rates due to a neutrino magnetic moment and to neutrino millicharge are proportional to the second power of these quantities. 

The expected signal spectrum results from the respective electron recoil spectrum in Figure~\ref{fig:exoticspec} being folded with the detection efficiency of the detector, which is a function of energy:
\begin{equation}
\frac{dN_{\rm tot}}{dE_{\rm recon}} = \int_{0}^{T_{max}} \frac{dN_{\rm tot}}{dT} \times S(T, E_{\rm recon}) dT,  
\label{eq:expectspec}
\end{equation} 
where $E_{\rm recon}$ is the reconstructed energy, $T_{max}=2E_{\nu}^{2}/(m_{e}+2E_{\nu})$ is the maximum recoil energy and $S(T, E_{\rm recon})$ is the signal efficiency of the data reduction steps as derived from Monte Carlo (MC) simulation. The signal efficiency curve for the millicharge analysis is shown in Figure~\ref{fig:milli_spec}; it corresponds to the function $S$ in Equation (\ref{eq:expectspec}). The green band reflects its systematic uncertainty. The spectra after the reduction process are shown in Figures~\ref{fig:milli_spec} and~\ref{fig:spec_BLNuMag}. We performed the detector simulation using the GEANT4 simulation package \cite{Geant4} for both signal and BG. The MC takes into account the non-linearity of the scintillation response in LXe as well as corrections derived from the detector calibrations. The electron equivalent energy is calculated from photoelectron counts (PE), with the conversion factor from PE to electron equivalent energy determined by comparing calibration data to MC simulation. Energy resolution is taken into account based on calibration data. Gaussian smearing is applied to MC to reproduce the data \cite{XMASS_detector}. The uncertainty of the scintillation efficiency coming from imperfect knowledge of the non-linearity of the scintillation response in LXe is included in the systematic uncertainty. 
The energy transferred in the interactions relevant to this paper ultimately becomes detectable as scintillation light emitted by electrons emerging from that interaction. As described in Section~\ref{sec:det}, radioactive sources were used to calibrate the detector response down to 1 keV. The uncertainty of this energy calibration is shown in Table~1 of \cite{XMASS_HP}. For lower energies, it is $\pm$4$\%$ at 1.65 keV and  $+$7$/-$4$\%$ at 1 keV. We conservatively assume that the scintillation efficiency below 1 keV is zero since we have a large uncertainty \cite{XMASS_subgevmod}.


\subsection{Dataset and event selection}
We analyzed the data, accumulated in the same period as \cite{XMASS_WIMP}, between November 2013 and March 2016. The total livetime is 711 days, which is slightly increased due to the recovery of some data in this analysis. 
The event-selection criteria were as follows: We required that (1) the ID trigger is not accompanied by an OD trigger, (2) there was no after pulse or Cherenkov event\footnote{The latter are primarily generated by $\beta$-rays from $^{40}$K in the PMT photocathodes.}, (3) $R(\rm Timing) <$ 38 cm, and (4) $R(\rm PE) <$ 20 cm, where $R(\rm Timing)$ and $R(\rm PE)$ were the distances from the center of the detector to the reconstructed vertex obtained by timing-based reconstruction \cite{XMASS_takeda} and by PE-based reconstruction \cite{XMASS_detector}, respectively. The $R(\rm Timing)$ selection is applied to suppress background events from the detector's inner surface. It was demonstrated that this selection is able to reduce events near the detector wall by a factor of ten around 5 keV as verified with the $^{241}$Am calibration source. The position resolution of $R(\rm PE)$ is about 4 cm around $R(\rm PE) =$ 20 cm at 5 keV. The fiducial mass of natural xenon in that 20 cm volume is 97 kg. 
The analyzed energy range was then set to be 2-15 keV for the neutrino millicharge search and 2-200 keV for the neutrino magnetic moment and dark photon searches. The analyzed energy range 2-200 keV covers the expected signal after applying all reduction steps; it contains about 98$\%$ of the signal MC events for neutrino-magnetic-moment interactions, $>$ 99$\%$ for dark photons of mass 1$\times$10$^{-3}$ MeV/$c^2$, and about 92$\%$ for dark photons of mass 10 MeV/$c^2$.

The systematic uncertainties in the signal were of two types. One came from the theoretical calculation of the signal. The uncertainty in the solar neutrino fluxes from the $pp$ and $^{7}$Be reactions are $\pm$0.6$\%$ and $\pm$7$\%$, respectively \cite{solarnu}. Also of this type is the uncertainty in the atomic effects in neutrino-electron interactions in xenon, which is $\pm$5$\%$ \cite{milli_estimate}. The other type of systematic uncertainty is related to the detector response. The most considerable systematic uncertainty in the signal is $\sim$15\% for the neutrino millicharge analysis, which came from the scintillation efficiency for electrons at low energy \cite{XMASS_subgevmod}. This is estimated by changing the scintillation efficiency parameters within the uncertainty obtained by calibration data with a $^{55}$Fe source \cite{XMASS_subgevmod}. For energies $>$ 30 keV, the uncertainty from the $R(\rm PE)$ cut, which corresponds to the uncertainty in fiducial mass, became dominant with $\sim$ 6\%. This was estimated from the difference of reconstructed position between data and MC in the $^{241}$Am and $^{57}$Co source calibrations. The uncertainty in the scintillation-decay time for electron recoils and in optical properties of the LXe were estimated in the same way as in \cite{XMASS_WIMP}. The uncertainty of signal strength due to the scintillation-decay time for electron recoils is $\sim$ 2 $\%$ for energies $<$ 10 keV and less than 1 $\%$ for energies $>$ 10 keV. The uncertainty of our signal estimates due to optical properties is $\sim$ 4 $\%$ for energies $<$ 10 keV and less than 1 $\%$ for energies $>$ 10 keV. 

\begin{figure}[t]
  \begin{center}
    \includegraphics[keepaspectratio=true,height=80mm]{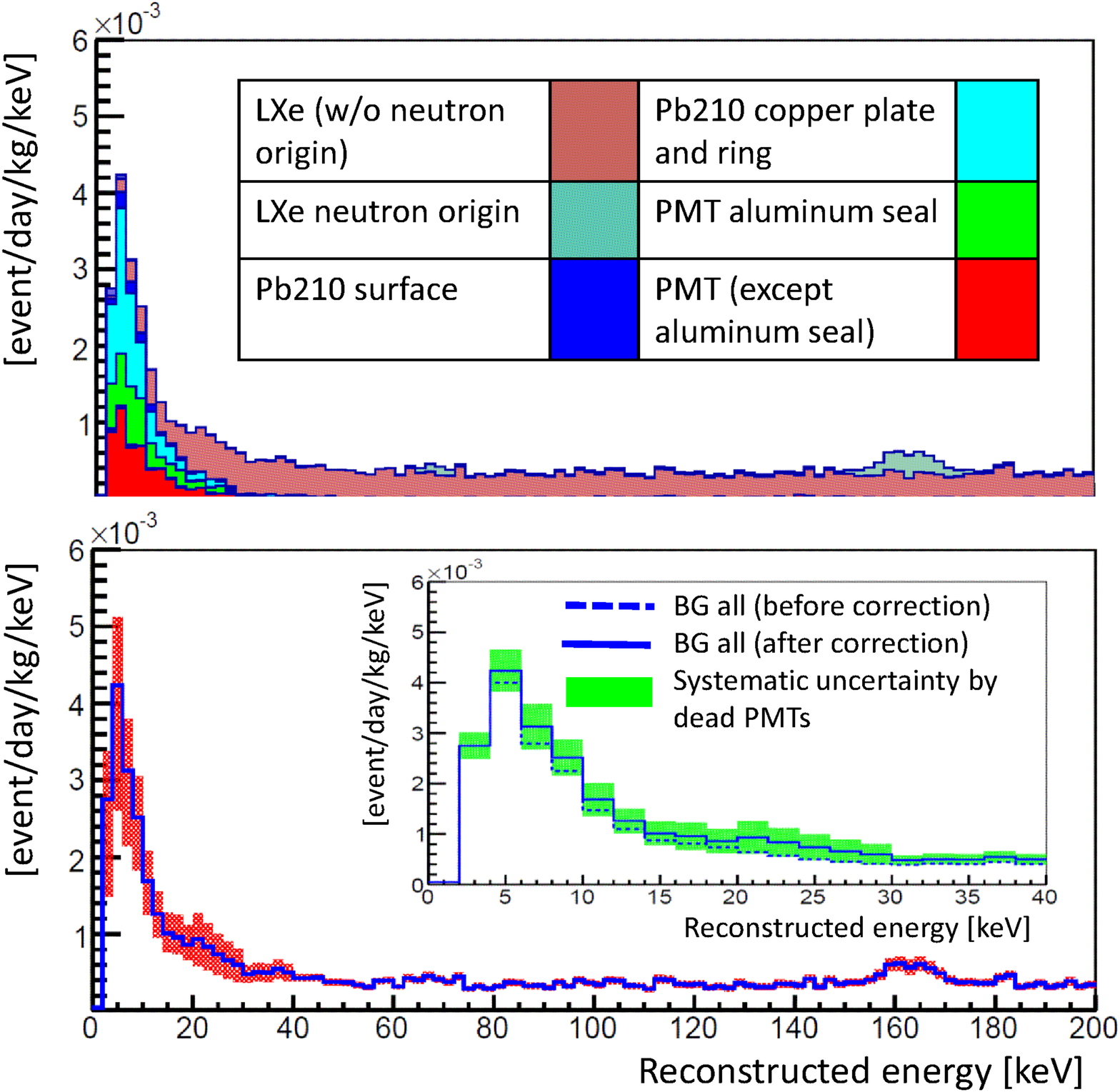}
  \end{center}
  \caption{The energy distribution of the BG simulation after event selection from 2 to 200 keV. The contributions to the BG originating from various types of events are indicated by the colored histograms in the top panel. The cumulative contribution of all the systematic errors is indicated by the red band in the bottom panel. The correction and the systematic uncertainty due to the correction of the dead PMTs in our BG model are shown in the inset of the bottom panel.}
\label{fig:spec_sys}
\end{figure}

Figure~\ref{fig:spec_sys} shows the energy distribution of the BG simulation from 2 to 200 keV after event selection.
The BG components in the fiducial volume were discussed in \cite{XMASS_WIMP} for $E_{\rm recon}<$ 30 keV and in \cite{XMASS_2nuECEC} for $E_{\rm recon}>$ 30 keV, respectively. The dominant BG component for $E_{\rm recon}<$ 30 keV derives from the radioactive isotopes (RI) that existed at the inner surface of the detector. The RI we took into account are $^{238}$U, $^{235}$U, $^{232}$Th, $^{40}$K, $^{60}$Co and $^{210}$Pb in the detector-surface materials, which include RI in the PMTs and copper plate and ring used for the PMT support structure. Moreover, the $^{210}$Pb accumulated on the inner surface of the detector is taken into account. RI induced surface events were often misidentified as events in the fiducial volume in the event reconstruction. All detector materials except for the LXe had been assayed using high-purity germanium (HPGe) detectors or a surface-alpha counter \cite{XMASS_copper}. The RI activities in the detector and their uncertainties were estimated by an analysis of alpha events and the energy spectrum without a fiducial volume cut.
The dominant BG component for $E_{\rm recon}>$30 keV was from RI dissolved in the LXe. Such events were distributed uniformly in the LXe and could not be removed by a fiducial-volume cut. Two categories of RI were found to be dissolved in the LXe: One was impurities such as $^{222}$Rn, $^{85}$Kr, $^{39}$Ar and $^{14}$C. The $^{222}$Rn and $^{85}$Kr activities were estimated using event coincidence in the full volume of the ID. In \cite{XMASS_2nuECEC}, we identified $^{39}$Ar and $^{14}$C in the detector from gas analysis of xenon samples and by performing spectral fitting. The other category were mostly xenon isotopes: $^{136}$Xe, which undergoes 2$\nu\beta\beta$ decay, and $^{125}$I, $^{131m}$Xe and $^{133}$Xe produced by neutron activation of common xenon isotopes. We estimated the concentration of $^{136}$Xe from its natural abundance and that of $^{125}$I from that of its precursor $^{124}$Xe and the thermal-neutron flux at the Kamioka Observatory, respectively. The concentrations of $^{131m}$Xe and $^{133}$Xe were estimated with a spectral fit performed in \cite{XMASS_2nuECEC}. The peak from $^{131m}$Xe can be seen near 160 keV in Figure~\ref{fig:spec_sys}.
We applied a data-driven correction to the simulated BG spectrum for $E_{\rm recon} <$ 40 keV in order to take into account the systematic difference in the mis-reconstruction rate caused by dead PMTs as we did in \cite{XMASS_HP}. The dead PMTs (9 out of 642 PMTs which had been found to be noisy or delivered strange responses) had been turned off. We evaluated the systematic difference of the probability with which events occurring close to the dead PMTs were reconstructed inside the fiducial volume. The difference between data and BG MC was found to be non-negligible below 40 keV. We applied a correction factor for the BG MC spectrum for such differences in each of the energy regions 2-5, 5-15, 15-20, 20-30 and 30-40 keV. These correction factors were estimated by comparing of the distance between the projection of the reconstructed vertex onto the detector surface and the dead-PMT position between data and BG MC in the fiducial volume. 
There are two systematic uncertainties associated with this correction factor. The first contribution was estimated by the difference in the correction factor estimated from the systematic difference of event rates in the fiducial volume by deliberately masking normal PMTs. The second contribution stems from the statistical uncertainty of the correction-factor estimate. The resultant correction and the systematic uncertainty of our BG model are shown in the inset of the bottom panel of Figure~\ref{fig:spec_sys}. These corrections amount to 0$\pm$10$\%$, 12$\pm$14$\%$, 14$\pm$19$\%$, 17$\pm$28$\%$, 46$\pm$34$\%$ and 23$\pm$20$\%$ in the energy regions 2-5, 5-15, 15-20, 20-30, and 30-40 keV, respectively.
The systematic uncertainties in the BG MC were basically the same as those used in our previous WIMP-search analysis \cite{XMASS_WIMP} for $E_{\rm recon}<$ 30 keV except for the dead PMT contribution. The dominant uncertainties came from uncertainty about irregular aspects of the geometry of e.g. gaps between the PMT holder and PMT bodies, and the surface roughness and the optical reflectivity of the PMT support structures. For 30-200 keV, we re-evaluate the systematic errors for uncertainties in the performance of the reconstruction, the scintillation-decay time, and the optical parameters of the LXe. Again most significant systematic uncertainly in this energy range comes from the position reconstruction, and is $\sim$6 \% as discussed before. Its estimation method was the same as for the signal MC.

    
\section{Search for exotic neutrino-electron interactions}
\label{sec:result}
\subsection{Fitting the energy spectrum}
Based on the BG estimate, we searched for the signatures of exotic neutrino-electron interactions by fitting the energy spectrum of the data with those of the BG MC and the respective signal MC. We define the fit by the following $\chi^{2}$:
\begin{equation}
\chi^{2} = \sum_{i}\frac{ ( D_{i} - B_{i} - \alpha \cdot S_{i} )^{2} }{ D_{i} + \sigma^{2}(B_{stat})_{i} + \alpha^{2} \cdot \sigma^{2}(S_{stat})_{i}} + \chi_{pull}^{2},\\
\end{equation}
 where $D_{i}$, $B_{i}$, and $S_{i}$ are the numbers of events in the data, the BG estimate, and the signal MC of the exotic neutrino interactions, respectively.  The index $i$ denotes the $i$-th energy bin. The value of $\alpha$ scales the signal-MC contribution. The quantity $B_{i}$ contains various kinds of BG sources. The terms $B_{i}$ and $S_{i}$ can be written as
\begin{equation}
B_{i}=\sum_{j}p_{j}(B_{ij}+\sum_{k}q_{k}\cdot\sigma(B_{sys})_{ijk}),\\
\end{equation}
\begin{equation}
S_{i}=S_{i}^{0}+\sum_{l}r_{l}\cdot\sigma(S_{sys})_{il},\\
\end{equation}
\begin{equation}
\chi^{2}_{pull}=\sum_{j}\frac{(1-p_{j})^{2}}{\sigma^{2}(B_{RI})_{j}}+\sum_{k}q_{k}^{2}+\sum_{l}r_{l}^{2}
\end{equation}
 where $j$ is the index of the BG components, and $k$, and $l$ are indices for systematic uncertainties in the BG and signal, respectively. We write the uncertainty in the amount of RI activity, systematic uncertainty in the BG and signal as $\sigma(B_{RI})_{j}$, $\sigma(B_{sys})_{ijk}$ and $\sigma(S_{sys})_{il}$, respectively. 
We scaled the RIs activities and the fraction of systematic errors by $p_{j}$, $q_{k}$ and $r_{l}$, respectively, while constraining them with a pull term ($\chi^{2}_{pull}$). 
The fitting range is 2-15 keV in the neutrino millicharge search, and is 2-200 keV in the dark photon and neutrino magnetic moment searches. We note that the constraints due to the RI activity from $^{14}$C, $^{39}$Ar, $^{131m}$Xe and $^{133}$Xe are not applied in the dark photon or neutrino magnetic moment searches because the expected signals are distributed at energies above 30 keV where spectrum fitting was performed to determine the RI activities in \cite{XMASS_2nuECEC}.
 
\subsection{Search for neutrino millicharge}
\begin{figure}[t]
  \begin{center}
   \includegraphics[keepaspectratio=true,height=70mm]{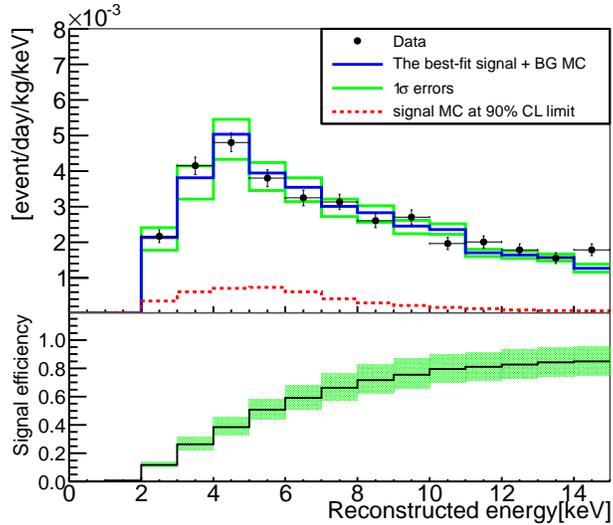}
  \end{center}
  \caption{(Top) The energy distribution after applying all cuts. The black points show the data. The blue histograms show the best-fit signal + BG MC simulation with 1 $\sigma$ errors shown by the green histograms. The red-dotted histograms show the 90\% CL upper limit for the neutrino-millicharge signal. (Bottom) The signal efficiency curve for the millicharge analysis. See text for detail.}
  \label{fig:milli_spec}
\end{figure}

\begin{figure}[tbp]
  \begin{center}
   \includegraphics[keepaspectratio=true,height=70mm,angle=-90]{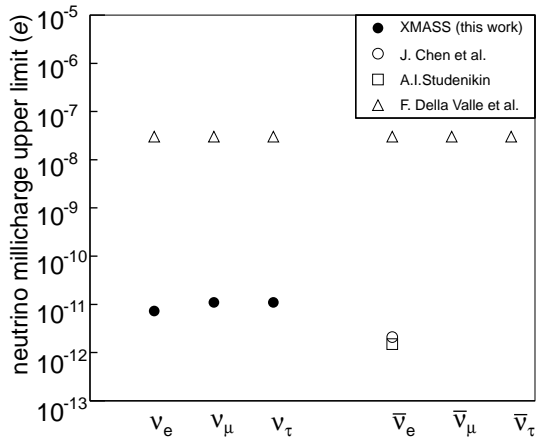}
  \end{center}
  \caption{90\% CL upper limits for neutrino millicharge for each flavor in ours and other experiments \cite{milli1, milli2, milli_PVLAS}. The limit from F. Della Valle {\it et al.} \cite{milli_PVLAS} depends on neutrino mass. It is for neutrino masses less than 10 meV.} 
  \label{fig:comp_milli}
\end{figure}

We found no significant signal excess, which would have been expected around 5 keV, and accordingly we set an upper limit for neutrino millicharge of $5.4 \times 10^{-12} e$ at the 90\% confidence level (CL), assuming all three species of neutrino have common millicharge. The best fit $\chi^2$ is obtained at zero millicharge. Figure \ref{fig:milli_spec} shows the data and the best-fit signal + BG MC with the signal MC at the 90\% CL upper limit. This limit is for neutrinos, not antineutrinos, and for neutrinos it is more stringent than the previous limit by more than three orders of magnitude \cite{milli_PVLAS}. Though the originally emitted solar neutrinos are $\nu_e$, the neutrinos arriving at Earth consist of all three flavors, which are produced by neutrino oscillations: At Earth 54$\pm$2\% are $\nu_e$, 23$\pm$1\% are $\nu_{\mu}$, and 23$\pm$1\% are $\nu_{\tau}$ \cite{PDG, koshio2}. Using this, we set upper limits for each flavor to be $7.3 \times 10^{-12} e$ for $\nu_e$, $1.1 \times 10^{-11} e$ for $\nu_{\mu}$, and $1.1 \times 10^{-11} e$ for $\nu_{\tau}$. These limits assume that only the neutrino flavor for which the limit is quoted carries a millicharge and thus contributes to the expected signal. Figure \ref{fig:comp_milli} compares our result with those of other experiments.

\subsection{Search for neutrino magnetic moment}
\begin{figure}[t]
  \begin{center}
    \includegraphics[keepaspectratio=true,height=65mm]{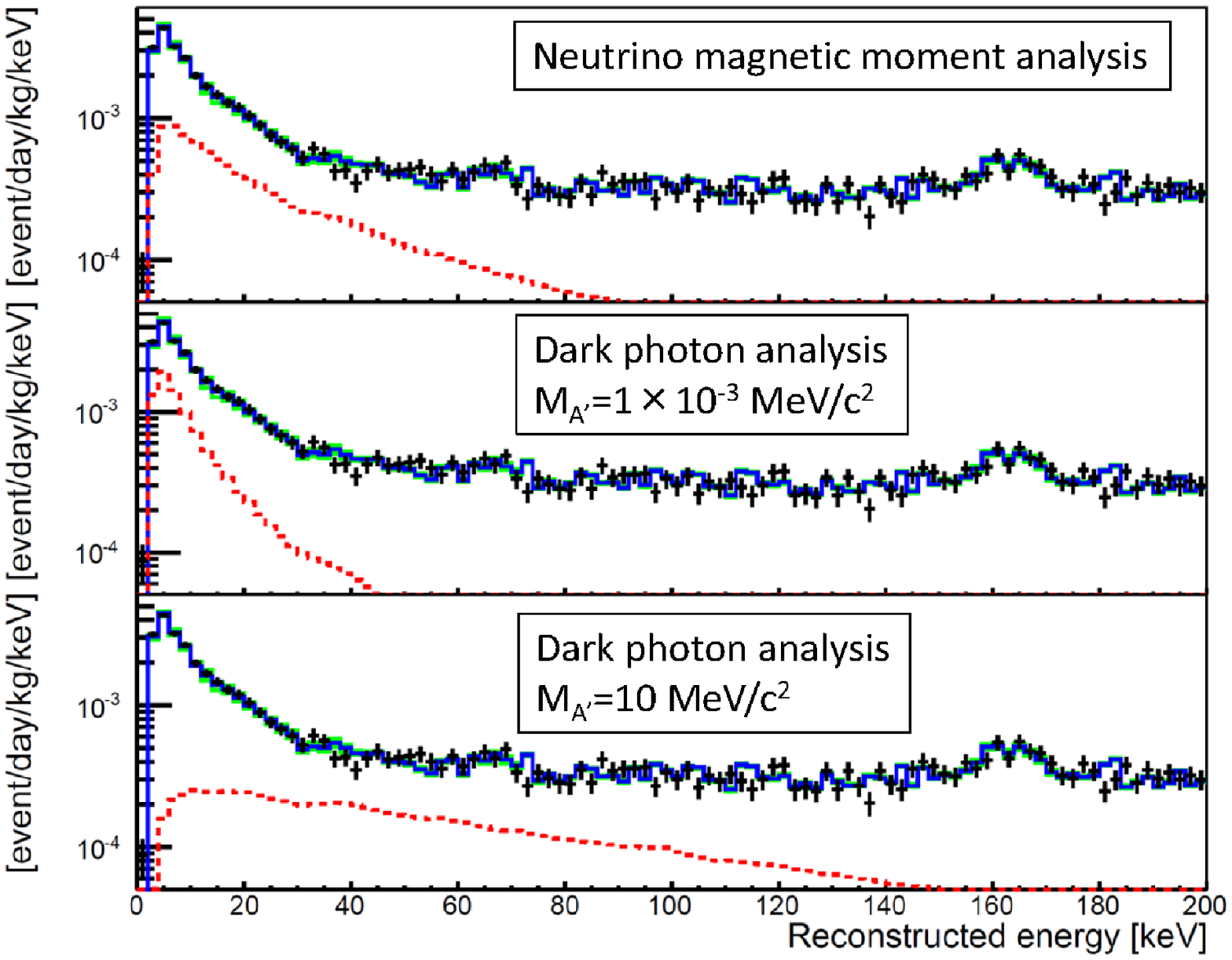}
  \end{center}
  \caption{The energy distribution of the data, the best fit signal + BG and the 90\% CL signal limit from 2 to 200 keV for the neutrino magnetic moment analysis (top) and the dark photon analysis (middle: dark photon mass $M_{A^{\prime}}$ = 1$\times 10^{-3}$ MeV/$c^2$, bottom $M_{A^{\prime}}$ = 10 MeV/$c^2$). The black points show the data. The blue histogram shows the signal + BG MC for the best fit with 1 $\sigma$ errors shown by the green histograms. The red-dotted histogram shows the 90\% CL upper limit for the signal. The peak near 160 keV stems from the decay of $^{131m}$Xe.}
  \label{fig:spec_BLNuMag}
\end{figure}

We also searched for a signal excess due to a neutrino magnetic moment, but again found no significant excess.
The top part of Figure~\ref{fig:spec_BLNuMag} shows the energy distribution of the data and the best-fit signal + BG. The contribution a neutrino magnetic moment at our 90\% CL signal limit would have made is also shown again. The best fit neutrino magnetic moment was $\mu_{\nu}=$1.3$\times$10$^{-10} \mu_{B}$, with a $\chi^{2} /d.o.f = 85.9/98$, while $\mu_{\nu} = 0$ yielded $\chi^{2} /d.o.f = 88.2/98$. The 90$\%$ CL upper limit for the neutrino magnetic moment is estimated from the $\chi^{2}$ probability density function to be $\mu_{\nu}=$1.8$\times$10$^{-10} \mu_{B}$.
 
\subsection{Search for neutrino interactions due to dark photons}
\begin{figure}[t]
  \begin{center}
   \includegraphics[keepaspectratio=true,height=80mm]{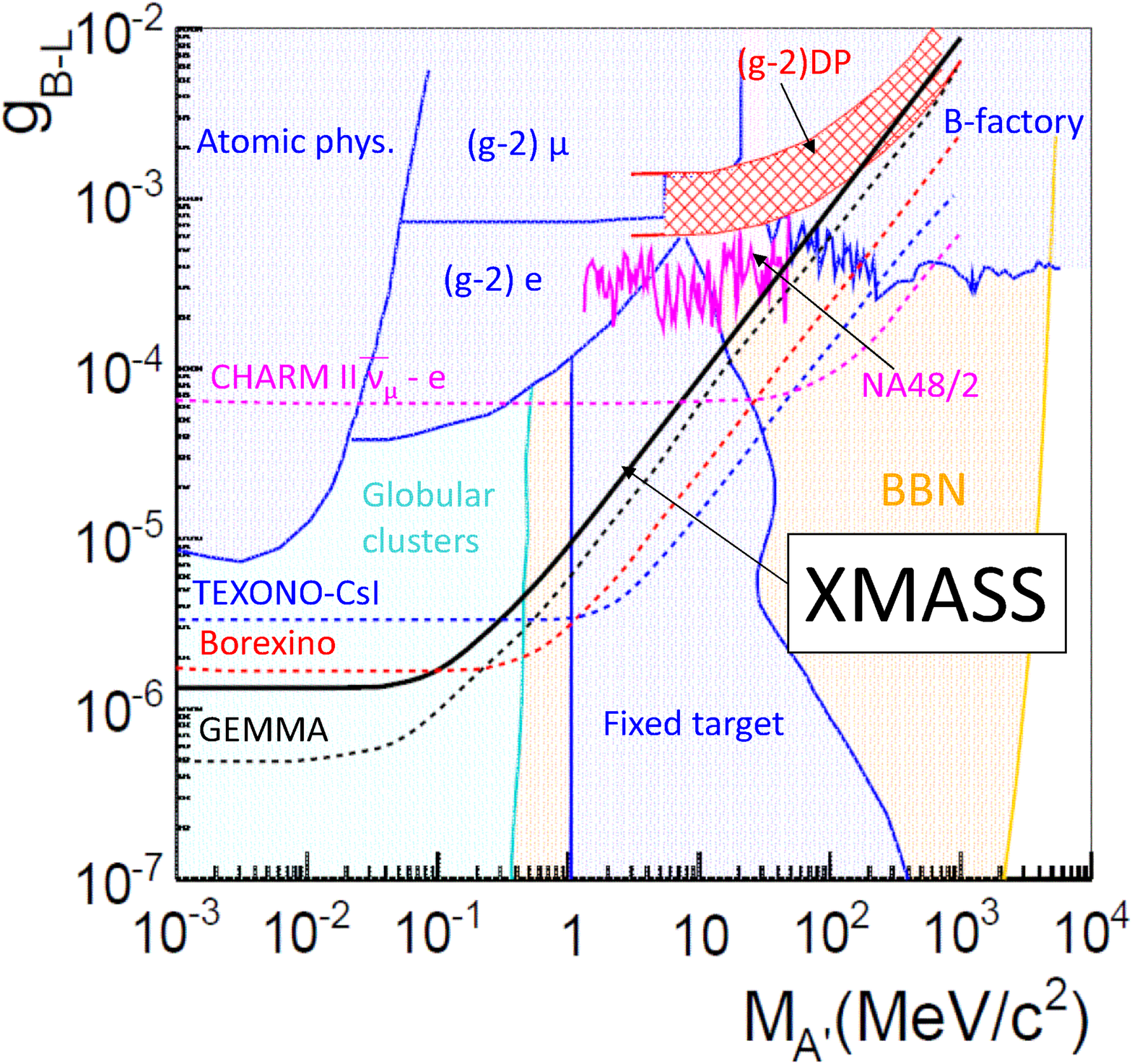}
  \end{center}
  \caption{90\% CL exclusion limits and allowed region on the coupling constant $g_{B-L}$ as a function of the dark photon mass $M_{A^{\prime}}$. The black-solid line shows the exclusion limit of our analysis (XMASS). The 2$\sigma$-allowed-region band from the muon $(g-2)$ experiment is shown as ``$(g-2)$ DP'' as the red-meshed region. The blue and magenta regions are excluded by laboratory experiments ($(g-2)_{\mu}$, $(g-2)_{e}$, atomic phys., fixed target, B-factory \cite{Bilmis} and NA48/2 \cite{NA48}), respectively. The cyan and orange regions are excluded by cosmological and astrophysical constraints (Globular clusters, BBN \cite{Bilmis}), respectively. BBN: the constraints of Big Bang nucleosynthesis on the mass of a light vector boson and its coupling constant to neutrinos in the B$-$L scenario. In this case, Dirac neutrinos $\nu_{R}$ are assumed \cite{Heeck}. 
The range of region follows as \cite{Bilmis}. The dotted lines are the estimated limit curves from neutrino-scattering experiments (GEMMA ($\bar{\nu}_{e}$), Borexino (solar $\nu$), TEXONO-CsI ($\bar{\nu}_{e}$) and CHARM II ($\bar{\nu}_{\mu}$)) from \cite{Bilmis}.}
  \label{fig:cont_BL}
\end{figure}

We also searched for a signal excess due to a dark photon with $M_{A^{\prime}}$ in the range from 1$\times 10^{-3}$  MeV$/c^{2}$ to 1$\times 10^{3}$ MeV$/c^{2}$. Again we found no significant excess.
The middle and bottom parts of Figure~\ref{fig:spec_BLNuMag} show the energy distributions of the data and the best-fit signal + BG. The contribution dark photons would have made at our 90\% CL limit is also shown in the figure. The value of $g_{B-L}$ from the best fit is $1.1\times10^{-6}$ with a $\chi^{2}/d.o.f = 85.3/98$ for $M_{A^{\prime}}=$1$\times 10^{-3}$  MeV$/c^{2}$ and is null with $\chi^{2}/d.o.f = 88.2/98$ for 10 MeV$/c^{2}$. The upper limits for $g_{B-L}$ for $M_{A^{\prime}}$=1$\times 10^{-3}$ MeV$/c^{2}$ and 10 MeV$/c^{2}$ are $1.3\times10^{-6}$ and $8.8\times10^{-5}$ at 90$\%$ CL, respectively. The 90$\%$ CL upper limit on the coupling constant as a function of the dark photon mass is shown in Figure~\ref{fig:cont_BL}, together with the limits and allowed region from other experimental and astrophysical analyses \cite{Bilmis}. 
Like the other neutrino and anti-neutrino scattering experiments we exclude a wide area in this parameter space, and for neutrinos our limit on $g_{B-L}$ is more stringent than Borexino's for $M_{A^{\prime}} <$ 0.1 MeV$/c^{2}$.
While the exclusion areas derived in \cite{Bilmis} from other experiments' publications already exclude an area larger than the one excluded by our analysis, our analysis is a dedicated one, incorporating our full knowledge of the detector response and our validated background models.
Also most of the parameter space for the $(g-2)$ dark photon prediction \cite{Bilmis} was excluded by our analysis. 

\section{Conclusions}
\label{sec:concl}
We conducted searches for exotic neutrino-electron interactions from solar neutrinos using 711 days of data in a 97 kg fiducial volume of the XMASS-I detector. We observed no significant signal. In the neutrino millicharge search, we set a neutrino millicharge upper limit of $5.4 \times 10^{-12} e$ at 90\% CL assuming all three species of neutrino have common millicharge. This is comparable to limits from previous experiments using antineutrinos. It is however three orders of magnitude better than the best previous limit for neutrinos \cite{milli_PVLAS}. We set upper limits for individual flavors at $7.3 \times 10^{-12} e$ for $\nu_e$, $1.1 \times 10^{-11} e$ for $\nu_{\mu}$, and $1.1 \times 10^{-11} e$ for $\nu_{\tau}$. Our upper limit for a neutrino magnetic moment is 1.8$\times$10$^{-10}\mu_{B}$.
Our result on dark photons in the $U(1)_{B-L}$ model imposes severe new restrictions on the coupling constant with neutrino from $M_{A^{\prime}}=$1$\times 10^{-3}$ to 1$\times 10^{3}$ MeV$/c^{2}$. 
In particular we almost exclude the area in which the $U(1)_{B-L}$ model can solve the $g-2$ anomaly.

\section*{Acknowledgments}
We thank Masahiro Ibe and Tsutomu Kakizaki for useful discussion about theoretical framework. Also we thank Yusuke Koshio for survival probability calculation for solar neutrino.
We gratefully acknowledge the cooperation of the Kamioka Mining and Smelting Company.
This work was supported by the Japanese Ministry of Education,
Culture, Sports, Science and Technology, the joint research program of the Institute for Cosmic Ray Research (ICRR), the University of Tokyo, 
Grant-in-Aid for Scientific Research, 
JSPS KAKENHI Grant No. 19GS0204, 26104004, and 19H05805
and partially by the National Research Foundation of Korea Grant funded
by the Korean Government (NRF-2011-220-C00006).




 
\end{document}